# Interactions between sub-10 nm iron and cerium oxide nanoparticles and 3T3 fibroblasts : the role of the coating and aggregation state

**M. Safi[1], H. Sarrouj[1], O. Sandre[2], N. Mignet[3] and J.-F. Berret[1][@]**

1 Matière et Systèmes Complexes, UMR 7057 CNRS Université Denis Diderot Paris-VII, Bâtiment Condorcet, 10 rue Alice Domon et Léonie Duquet, 75205 Paris, France
2 UPMC Univ Paris VI – Laboratoire de Physico-chimie des Electrolytes, Colloïdes et Sciences Analytiques UMR 7195 UPMC Univ Paris 6 / CNRS / ESPCI Paristech, 4 place Jussieu, 75252 Paris Cedex 05 (France)
3 CNRS UMR8151, Faculté de Pharmacie, 4 avenue de l'Observatoire, 75270 Paris, France

**Abstract :** Recent nanotoxicity studies revealed that the physico-chemical characteristics of engineered nanomaterials play an important role in the interactions with living cells. Here, we report on the toxicity and uptake of the cerium and iron oxide sub-10 nm nanoparticles by NIH/3T3 mouse fibroblasts. Coating strategies include low-molecular weight ligands (citric acid) and polymers (poly(acrylic acid), $M_W$ = 2000 g mol$^{-1}$). Electrostatically adsorbed on the surfaces, the organic moieties provide a negatively charged coating in physiological conditions. We find that most particles were biocompatible, as exposed cells remained 100% viable relative to controls. Only the bare and the citrate-coated nanoceria exhibit a slight decrease of the mitochondrial activity at very high cerium concentrations (> 1 g L$^{-1}$). We also observe that the citrate-coated particles are internalized/adsorbed by the cells in large amounts, typically 250 pg per cell after a 24 h incubation for iron oxide. In contrast, the polymer-coated particles are taken up at much lower rates (< 30 pg per cell). The strong uptake shown by the citrated particles is related to the destabilization of the dispersions in the cell culture medium and their sedimentation down to the cell membranes. In conclusion, we show that the uptake of nanomaterials by living cells depends on the coating of the particles and on its ability to preserve the colloidal nature of the dispersions.

# I – Introduction

During the last years, engineered inorganic particles have emerged as fundamental constituents in the development of nanotechnology. Engineered nanoparticles are ultrafine colloids of nanometer dimensions with highly ordered crystallographic structures. These particles exhibit usually remarkable electronic, magnetic or optical properties that can be exploited in a variety of applications. At the nanometer scale, some of these nanocrystals exhibit physical properties that are dramatically different from those of their bulk materials. Gold, iron oxide and quantum dots belong to this class [1-3]. For some others, the interesting attributes are not uniquely related to the nanometer size, but depend on the nature and crystallinity of the particles. This is the case for titanium, zinc or cerium oxides which are used industrially for their ability to filter the electromagnetic waves in the UV range. These nanoparticles have therefore found applications in cosmetics, coating and surface treatment. Details of the synthesis and applications of engineered inorganic particles are available in several reviews [2, 4].

In contrast to conventional chemicals, the possible risks of using nanomaterials for human health and the environment have not been yet fully evaluated [5-7]. To estimate these risks, large research efforts were directed towards the development of toxicology assays. The objectives of these assays, such as the MTT, neutral red or WST1 are the quantitative determination of the viability of living cells that were incubated with nanomaterials [5, 6]. In recent years, the viability of many different cell lines was investigated with respect to a wide variety of engineered nanoparticles. Recent reviews attempted to recapitulate the main features of the toxicity induced by nanomaterials. One of these features deals with the coating of the particles. In most *in vitro* studies, the chemistry of the interfaces between the nanocrystals and the solvent was anticipated to be a key feature of cell-nanoparticle interactions.

Although the toxicity of bare uncoated particles was also examined [8-14], the majority of engineered nanomaterials studied on cell lines were first coated with ligands, peptides or macromolecules. Ligands such as citric acid [15] and dimercaptosuccinic acid (DMSA) [15-18] were generally adsorbed at the surface of the metal oxide particles through electrostatic interactions. Thanks to the carboxylate groups borne by these ligands, the particles became charged and exhibited an enhanced stability in aqueous media at physiological pH. Auffan and coworkers reported that DMSA coated iron oxide nanoparticles up to a concentration of 0.1 mg mL$^{-1}$ produced weak toxicity and no genotoxic effects in human fibroblasts [16]. Interestingly, these authors also observed that in the cell medium, DMSA-coated particles agglomerated due to the formation of disulfur bridges, indicating a marginal stability of these particles in complex environments. The other class of coating examined in these studies utilized polymeric chains such as dextran [19-21], poly(ethylene glycol) (PEG) or PEG based systems [22-25] and poly(vinyl alcohol) (PVA) [26-28]. More complicated architectures such as block or graft copolymers were also attempted [24, 29-31]. Attached on the particle surfaces through covalent or ionic binding, these hydrosoluble polymers usually formed a corona around the inorganic cores. The role of the corona was to enhance the colloidal stability in complex environments, as well as to induce an anti-fouling effect with respect to the serum proteins. Villanueva *et al.* compared the effects of positive, neutral and negative polymer coatings on magnetite nanoparticles on the viability of HeLa cells [21]. These authors have found an overall low toxicity of coated nanomaterials towards living cells, but an effective and enhanced uptake of the positively charged particles. Other reports underlined that not only the electrostatic charges of the corona was important but also its radial extension [20, 24, 25]. In terms of uptake, most work dealing with the cell-nanoparticle interactions revealed that the particles were internalized by the cells, sometimes in large quantities.

In the present paper, we investigated the *in vitro* toxicity and internalization of sub-10 nm cerium (nanoceria, $CeO_2$) and iron oxide (maghemite, $\gamma$-$Fe_2O_3$) nanoparticles using mice NIH/3T3 fibroblasts. Nanoceria and maghemite were selected because the particles have shown promising features for biomedical and industrial applications, and it is foreseen that their importance will increase in future technological developments [5, 12, 32, 33]. Both particles were synthesized using "soft chemistry" routes that provided nanocolloids of different nature and crystal structures, but with the same physico-chemical features. This analogy allowed us to apply the same protocols for their coating, resulting in a unique opportunity to compare the influence of the nature of the inorganic materials to that of the coating in toxicity assays. Recently, we have developed an easy and widely applicable method to adsorb ion-containing polymers (in this case poly(acrylic acid) with molecular weight 2000 g mol$^{-1}$) onto the nanoparticle surfaces [34, 35]. We have found that this low-molecular weight polymers augmented the hydrodynamic diameters of the particles by only 4 nm, and at the same time preserved the long term colloidal stability in most water based solvents, including buffers and cell culture media [36]. This noticeable increase in stability as compared to classical ligand-coated particles has prompted us to perform toxicity assays, and to explore the effect of the dispersion state on intracellular uptake.

# II – Materials and Methods

## II.1 – Nanoparticles and coating

The synthesis of the cerium and iron oxide nanoparticles used the technique of « soft chemistry » based on the polycondensation of metallic salts in alkaline aqueous media. The synthesis has been previously described, and we refer to this work for more details [34, 37]. Images of transmission electron microscopy obtained from the cerium and iron oxide dispersions at weight concentration c = 0.2 wt. % are illustrated in Fig. 1a and Fig. 1b respectively. The nanoceria consisted of isotropic agglomerates of 2 nm-crystallites with faceted morphologies, whereas the iron oxide nanoparticles exhibited more compact and spherical structures. An analysis of these and other TEM photographs allowed us to derive the size distributions of the particles. In both instances, they were found to be well-accounted for by a log-normal function, with median diameter $D_0$ and polydispersity s. Here, the polydispersity was defined as the ratio between the standard deviation $\sqrt{\langle D^2 \rangle - \langle D \rangle^2}$ and the average diameter, where <...> denotes the average. In the present study, one nanoceria and two maghemite batches were synthesized with median diameters in the range of 7 - 8 nm and a polydispersity between 0.15 and 0.30. $D_0$ and s for bare particles are listed in Table I. During the synthesis, the particles were positively charged, with nitrate counterions adsorbed on their surfaces.

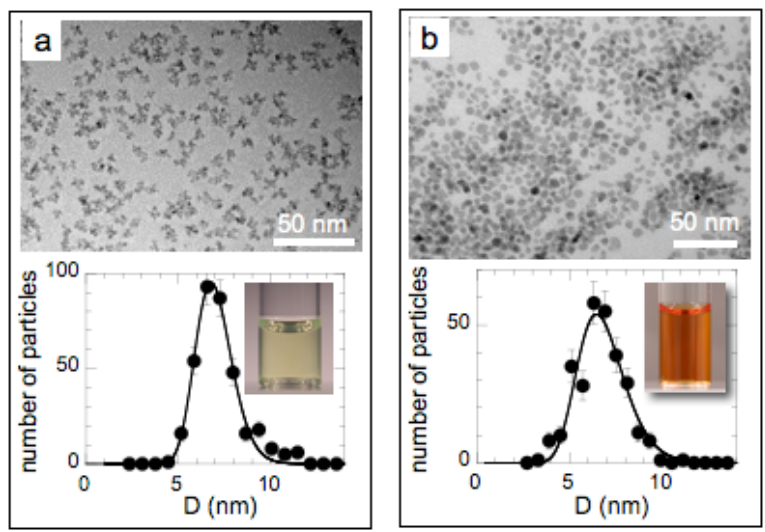

*Figure 1 :* *Transmission electron microscopy images (upper panels) of the nanoparticles investigated in this work : a) cerium oxide (CeO$_2$), b) iron oxide (γ-Fe$_2$O$_3$). A statistical treatment of the TEM data allowed us to derive the size distributions of the different nanocrystals (lower panels). The TEM images presented here were recorded with a magnification ×120000. The insets show photographs of vials containing the cerium and iron oxide dispersions.*

The resulting electrostatic repulsion between nanocolloids insured stability of the dispersions over several years. Using dynamic light scattering, the molecular weight and the hydrodynamic diameter of the particles in acidic conditions were determined (Table I). Note that the hydrodynamic sizes in Table I appeared larger than the median diameters found by TEM. For γ-Fe$_2$O$_3$, these differences were attributed to the slight anisotropy of the particles (aspect ratio 1.2) and to their intrinsic polydispersity [35]. The characterization of CeO$_2$ and γ-Fe$_2$O$_3$ nanocrystals in terms of crystal structure, electrophoretic mobility and light absorbance is described in Supporting Information. For this work, we used two different coating agents : citric acid molecules and poly(acrylic acid) polymers.

*Citric acid* : Citric acid is a weak tri-acid of molecular weight $M_W$ = 192.1 g mol$^{-1}$, which has three acidity constants at $pK_{A1}$ = 3.1, $pK_{A2}$ = 4.8 and $pK_{A3}$ = 6.4. For iron oxide, the complexation of the surface charges with citric acid (Sigma Aldrich) was performed during the synthesis by adding tri-sodium citrate in excess under vigorous stirring, followed by washing steps with acetone and diethyl ether.

| nanoparticles | $D_0$ nm | s | $M_W$ kg mol$^{-1}$ | $D_H$ nm |
|---|---|---|---|---|
| CeO$_2$ | 7.0 | 0.15 | 330 | 9 |
| γ-Fe$_2$O$_3$ (acidic) | 7.1 | 0.26 | 1400 | 14 |
| γ-Fe$_2$O$_3$ (citrated) | 8.5 | 0.29 | 3000 | 23 |

**Table I** : Median diameter ($D_0$), polydispersity (s), molecular weight ($M_W$) and hydrodynamic diameter $D_H$ of the bare nanoparticles studied in this work. The size distributions of the particles were assumed to be log-normal (Eq. 1) for the three batches.

| Coated NPs | $D_H$ nm | h nm |
|---|---|---|
| Cit–CeO$_2$ | 9 | < 0.5 |
| PAA$_{2K}$–CeO$_2$ | 13 | 2 |
| Cit–γ-Fe$_2$O$_3$ | 23 | < 0.5 |
| PAA$_{2K}$–γ-Fe$_2$O$_3$ | 19 | 2.5 |

**Table II** : Hydrodynamic diameters $D_H$ and adlayer thicknesses h derived from dynamic light scattering measurements for the coated nanoparticles dispersed in de-ionized water.

It allowed to reverse the surface charge of the particles from cationic at low pH to anionic at high pH, through a ionization of the carboxyl groups. At pH 8, the particles were stabilized by electrostatic interactions [34]. For nanoceria, the adsorption of citrate ions was performed after the synthesis. Citrate ions were characterized by adsorption isotherms, *i.e.* the adsorbed species were in equilibrium with free citrates molecules dispersed in the bulk. The concentration of free citrates in the bulk was kept at the value of 8 mM [38, 39] both in water and in culture medium. It should be noticed that the hydrodynamic diameter of the bare and citrated particles were identical within the experimental accuracy, indicating a layer thickness less than 1 nm (Table II). The citrate-coated particles are denoted Cit–CeO$_2$ and Cit–γ-Fe$_2$O$_3$ in the sequel of the paper.

*Poly(acrylic acid)* : During the last years, poly(acrylic acid) was frequently used as a coating agent of inorganic particles [32, 34, 40-43]. Poly(sodium acrylate), the salt form of poly(acrylic acid) with a molecular weight $M_W = 2000$ g mol$^{-1}$ and a polydispersity of 1.7 was purchased from Sigma Aldrich and used without further purification. It is denoted here as PAA$_{2K}$. In order to adsorb polyelectrolytes on the surface of the nanoparticles, we followed the precipitation-redispersion protocol, as described elsewhere [34, 35]. The precipitation of the cationic cerium or iron oxide dispersions by PAA$_{2K}$ was performed in acidic conditions (pH 2). The precipitate was separated from the solution by centrifugation, and its pH was increased by addition of ammonium hydroxide. The precipitate redispersed spontaneously at pH 7 - 8, yielding a clear solution that then contained the polymer coated particles. This simple technique allowed to produce large quantities of coated particles (> 1 g of oxides) within a relatively short time (< 1 h). The hydrodynamic sizes of PAA$_{2K}$–CeO$_2$, PAA$_{2K}$–$\gamma$-Fe$_2$O$_3$ were found to be $D_H$ = 13 and 19 nm, respectively. These values were 4 - 5 nm larger than the hydrodynamic diameter of the uncoated particles, indicating a corona thickness $h$ = 2 - 2.5 nm (Table II). In terms of coverage, the number of adsorbed chains per particle was estimated to be 50 for CeO$_2$ [34] and 180 for $\gamma$-Fe$_2$O$_3$. As for the citrated particles, it was checked by electrokinetic measurements that the PAA$_{2K}$ coating resulted in strongly anionic charged interfaces [29, 34, 42]. Values of the electrophoretic mobilities were found at $\mu_E$ = -1.87×10$^{-4}$, -3.35×10$^{-4}$, -3.76×10$^{-4}$ and -3.52×10$^{-4}$ cm$^2$ V$^{-1}$ for Cit–CeO$_2$, PAA$_{2K}$–CeO$_2$, Cit–$\gamma$-Fe$_2$O$_3$ and PAA$_{2K}$–$\gamma$-Fe$_2$O$_3$ respectively (see Supporting Information).

As a final step of the procedures described above, the dispersions were dialyzed against DI-water which pH was first adjusted to 8 by addition of sodium hydroxide (Spectra Por 2 dialysis membrane with MWCO 12 kD). For the citrate-coated particles, DI-water was supplemented with 8 mM of free citrates. At this pH, 90 % of the carboxylate groups of the citrate and PAA$_{2K}$ coating were ionized. The suspension pH was adjusted with reagent-grade nitric acid (HNO$_3$) and with sodium or ammonium hydroxides. For the assessment of the stability with respect to ionic strength, sodium and ammonium chloride (NaCl and NH$_4$Cl, Fluka) were used in the range $I_S$ = 0 – 1 M [36].

## II.2 – Experimental techniques

*Transmission electron microscopy (TEM)* : TEM experiments were carried out on a Jeol-100 CX microscope at the SIARE facility of University Pierre et Marie Curie (UPMC). The TEM images of the cerium and iron oxide nanoparticles (magnification ×120000) were analyzed using the ImageJ software (http://rsb.info.nih.gov/ij/). The diameter and polydispersity for CeO$_2$ were in good agreement with those determined by cryogenic transmission electron microscopy in an earlier report [42].

*Dynamic Light Scattering* : Dynamic light scattering was performed on a Brookhaven spectrometer (BI-9000AT autocorrelator, $\lambda$ = 632.8 nm) for measurements of the Rayleigh ratio $\mathcal{R}(q,c)$ and the collective diffusion constant D(c). In dynamic light scattering, the collective diffusion coefficient D was determined from the second-order autocorrelation function of the scattered light. From the value of the coefficient, the hydrodynamic diameter of the colloids was calculated according to the Stokes-Einstein relation, $D_H = k_BT/3\pi\eta_S D$, where $k_B$ is the Boltzmann constant, T the temperature (T = 298 K) and $\eta_0$ the solvent viscosity ($\eta_S$ = 0.89×10$^{-3}$ Pa s for water and 0.95×10$^{-3}$ Pa s for DMEM supplemented with calf serum and at T = 25 °C). The autocorrelation functions were interpreted using the cumulants and the CONTIN fitting procedure provided by the instrument software. In the present work, the hydrodynamic diameters and the intensity scattered by nanoparticles dispersed in a cell growth medium were evaluated as a function of time and concentration [36]. In case of a destabilization of the sol, the scattering intensity was expected to grow rapidly, as well as the hydrodynamic diameter. With an accuracy better than 5 % on the intensity and 10 % on the diameter, this technique was very sensitive to the dispersion state.

*UV-visible spectrometry* : A UV-visible spectrometer (SmartSpecPlus from BioRad) was used to measure the absorbance of bare and coated nanoparticles dispersion in water. In the range $\lambda$ = 200 – 800 nm, the absorbance was related to the nanoparticle concentration by the Beer-Lambert law :

$$\text{Abs}(c,\lambda) = -\log T(c,\lambda) = \varepsilon(\lambda)\ell c \qquad (1)$$

where $\ell$ (= 1 cm) is the optical path length, c the nanoparticle concentration, $\varepsilon$ the molar absorption coefficient (cm$^{-1}$ mM$^{-1}$) and T the transmission. Taking advantage of the strong absorbance of the cerium and iron oxides below 400 nm, the nanoparticle concentrations could be determined very accurately, with an uncertainty better than 5×10$^{-4}$ wt. % (5×10$^{-2}$ mM). The molar absorption coefficients $\varepsilon$ for uncoated $CeO_2$ and $\gamma$-$Fe_2O_3$ are shown in Supporting Information (SI.4). It was also verified that the citrate and polymer coating did not modify the absorption coefficient. Data in Figs. SI.4 were used to estimate the amount of particles taken up by the cells as a function of the time. Aliquots of the supernatant located above the NIH/3T3 were collected at different times after the incubation of the cells. The experiments were made in duplicate for the concentrations {[Ce],[Fe]} = 1 and 10 mM. In some cases, the particles precipitated in the cell culture medium, increasing their absorbance properties. The aliquots were thus diluted, so as to minimize the $\lambda^{-1}$-contribution of large aggregates [44]. In these cases, the accuracy in the oxide concentration increased to ± 3 %.

*Optical microscopy* : Phase-contrast images of the cells were acquired on an IX71 inverted microscope (Olympus) equipped with 10× and 60× objectives.

Data acquisition and treatment were monitored with a Photometrics Cascade camera (Roper Scientific), Metaview (Universal Imaging Inc.) and ImageJ softwares.

### II.3 – Cell culture and MTT Assays

NIH/3T3 fibroblast cells from mice were grown as a monolayer in Dulbecco's Modified Eagle's Medium (DMEM) with High Glucose (4.5 g $L^{-1}$) and stable Glutamine (PAA Laboratories GmbH, Austria). This medium was supplemented with 10% Fetal Bovine Serum (FBS), and 1% penicillin/streptomycin (PAA Laboratories GmbH, Austria), referred to as cell culture medium. Exponentially growing cultures were maintained in a humidified atmosphere of 5% $CO_2$ - 95% air at 37°C, and under these conditions the plating efficiency was 70 – 90 % and the doubling time was 12 – 14 h. Cell cultures were passaged once or twice weekly using trypsin-EDTA (PAA Laboratories GmbH, Austria) to detach the cells from their culture flasks and wells. Subconfluent cells (50 – 60 %) were detached using a trypsin-EDTA solution followed by the addition of complete cell culture medium (DMEM) to neutralize trypsin.

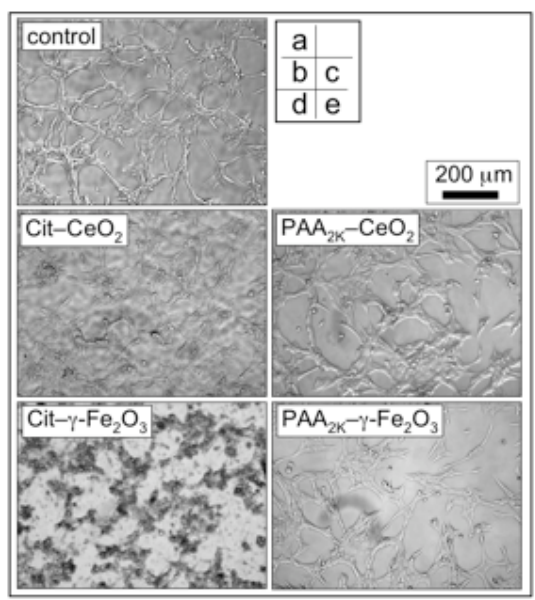

***Figure 2*** *: Transmission optical microscopy (10×) images of NIH/3T3 fibroblasts that were grown without particles (2a). In 2b, 2c, 2d and 2e, the cells were incubated with Cit–CeO₂, PAA₂ₖ–CeO₂, Cit–γ-Fe₂O₃ and PAA₂ₖ–γ-Fe₂O₃ nanoparticles during 24 h, the metal concentration being {[Ce],[Fe]} = 1 mM. For contrast reasons, the supernatant containing the citrated particles (2b and 2d) was removed and replaced by pristine medium.*

The cells were pelleted by centrifugation at 1200 rpm for 5 min at 26°C. Supernatants were removed, and cell pellets were re-suspended in assay medium and counted using a hemocytometer. For cell counting, the fibroblasts were seeded in T25-flasks. The surface occupied by the cells as a function of the time was estimated by taking microscopy images (objective 10×) at regular time intervals between 0 and 96 h. The cell density was then calculated using an ImageJ statistical analysis of different fields of view, assuming a constant value for the cell area (400 $\mu m^2$/cell). MTT assays were performed with both coated and uncoated cerium and iron oxide nanoparticles for metal molar concentrations {[Ce], [Fe]} between 10 µM to 10 or 50 mM. Cells were seeded into 96-well microplates, and the plates were placed in an incubator overnight to allow for attachment and recovery. Cell densities were adjusted to $2\times10^4$ cells per well (200 µl). After 24 h, the nanoparticles were applied directly to each well using a multichannel pipette to triplicate culture wells, and cultures were incubated for 24 h at 37°C. The MTT assay depends on the cellular reduction of MTT (3-(4,5-dimethylthiazol-2-yl)-2,5-diphenyl tetrazolium bromide, Sigma-Aldrich Chemical) by the mitochondrial dehydrogenase of viable cells forming a blue formazan product which can be measured spectrophotometrically [45]. MTT was prepared at 5 mg mL$^{-1}$ in PBS (with calcium and magnesium, Dulbecco's, PAA Laboratories) and then diluted 1 to 5 in medium without serum and without Phenol Red. After 24 h of incubation with nanoparticles, the medium was removed and 200 µl of the MTT solution was added to the microculture wells. After 4 h incubation at 37°C, the MTT solution was removed and 100 µl of 100% DMSO were added to each well to solubilize the MTT-formazan product. The absorbance at 562 nm was then measured with a microplate reader (Perkin-Elmer). Prior to the microplate UV-Vis spectrometry, MTT assays without particles were carried out with cell populations ranging from 5000 to 500000 cells and it was checked that the absorbance of DMSO solutions at 562 nm was proportional to the initial number of cells.

# III – Results and Discussion

## III.1 - Cell counting

In order to determine their optimal growth conditions, the fibroblasts were first plated in culture medium without particles. Fig. 2a provides an illustration of the NIH/3T3 observed by optical microscopy at a 50 % coverage (objective 10×). Figs. 2b, 2c, 2d and 2e display NIH/3T3 fibroblasts that were exposed during 24 h to Cit–$CeO_2$, $PAA_{2K}$–$CeO_2$, Cit–γ-$Fe_2O_3$ and $PAA_{2K}$–γ-$Fe_2O_3$ nanoparticles respectively, at a concentration {[Ce],[Fe]} = 1 mM. Note that for contrast reasons the supernatant containing the citrated particles was removed and after thorough washing with PBS, it was replaced by pristine medium. For the PAA$_{2K}$-coated particles, the images were recorded in the same conditions as for the control, the particles being dispersed in the cell medium. Fig. 2 revealed first that the fibroblasts have approximately the same number density after a 24 h incubation, indicating that in the 5 wells the cells grew at the same rate. A quantitative analysis of the cell populations will evidence small discrepancies between the different cultures. In Fig. 2 however, there is a marked difference between cells incubated with citrate and with PAA$_{2K}$-coated particles. Due to a massive internalization and/or adsorption of the nanomaterial by the cells, the cells exposed to the citrate-coated particles were more difficult to detect.

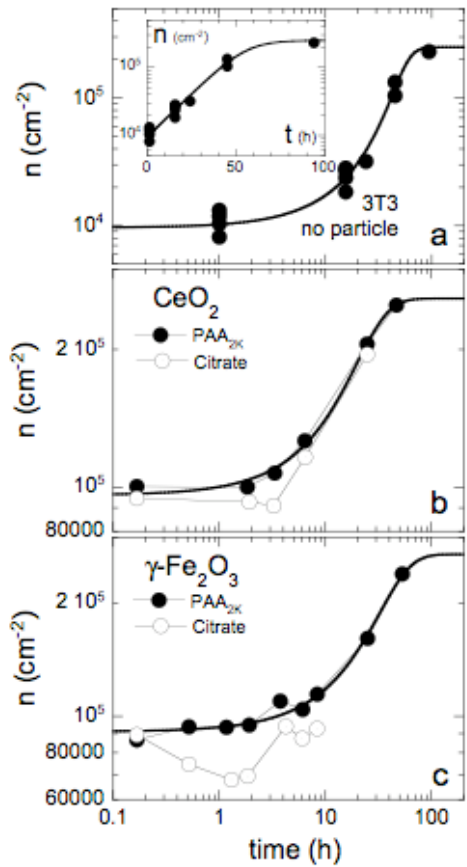

*Figure 3 :* *Number densities of NIH/3T3 fibroblasts as a function of time for different experimental conditions. In a), the cells were grown without nanoparticle added, whereas in b) and c) the NIH/3T3 were incubated with Cit–CeO$_2$, PAA$_{2K}$–CeO$_2$, Cit–γ-Fe$_2$O$_3$ and PAA$_{2K}$–γ-Fe$_2$O$_3$ nanoparticles at concentrations {[Ce], [Fe]} = 10 mM. The continuous lines were obtained from best fit calculations using Eq. 5, and adjustable parameters given in the text. This procedure allowed an accurate determination of the duplication time τ$_D$. Inset in a) : same data as in the main frame but plotted in the semilogarithmic representation in order to emphasize the exponential growth at short times.*

The dark patterns seen in the bottom left image were stemming from internalized or adsorbed Cit–$\gamma$-Fe$_2$O$_3$ nanoparticles. In contrast, the cells incubated with the PAA$_{2K}$–CeO$_2$ and with PAA$_{2K}$–$\gamma$-Fe$_2$O$_3$ behaved as the control. No aggregate of particles could be detected even at a higher magnification (60×). The values of the cell coverage were obtained through an ImageJ statistical analysis of the surface occupied by the cells with respect to the overall field of observation. The range of coverage investigated was comprised between 5 % to 100 % (confluence). An average surface of 400 $\mu m^2$ for the NIH/3T3 was measured independently on a panel of 50 cells. This value was used to translate the surface coverage into cell density. Fig. 3a displays the number density of NIH/3T3 cells as a function of time in a double logarithmic representation (main frame) and in semilogarithmic representation (inset). Without nanoparticles, the cell population exhibited an exponential growth over the first 48 h (recognized by the straight line in the inset, t < 48 h), and then a saturation at a value $n_S = 2.5 \times 10^5$ cm$^{-2}$. The exponential increase of the cell density was adjusted using the expression :

$$n(t) = n_0 \, 2^{t/\tau_D} \tag{2}$$

where $n_0$ denotes the initial cell density and $\tau_D$ the duplication time. The double logarithmic representation in the main frame of Fig. 3a aimed to emphasize the cell proliferation at short times (see below). In order to describe the time evolution of the cell populations over the whole time range, a modified exponential growth model was developed. This model took in to account the slowing-down of the growth as the coverage of the substrate reached saturation. The continuous lines in Fig. 3a resulted from best fit calculations using the model prediction of the form :

$$n(t) = n_0 \, 2^{t/\tau_D} \left( 1 + \frac{n_0^2}{n_S^2} \left( 2^{2t/\tau_D} - 1 \right) \right)^{-1/2} \tag{3}$$

where $n_S$ is the final cell density. Derived in Supporting Information (Eq. SI.7), this expression allowed an accurate determination of the duplication time $\tau_D$. The continuous lines in Figs 3a were obtained using $n_0 = 9.6 \times 10^3$ cm$^{-2}$ cells and $\tau_D = 11.9$ h.

Figs. 3b and 3c display the results obtained on cells incubated with CeO$_2$ and $\gamma$-Fe$_2$O$_3$ nanoparticles at {[Ce,Fe]} = 1 mM. The growth laws for cells incubated with the PAA$_{2K}$-coated nanoparticles (closed symbols) exhibit similar behaviors as in experiments without nanoparticles : n(t) evolved slowly up to 10 hours and then increased up to the saturation level $n_S$. The continuous lines in Figs. 3b and 3c were computed from Eq. 5 using $n_0 = 9.6 \times 10^4$ cm$^{-2}$ and $9.1 \times 10^4$ cells per

cm$^{-2}$, $\tau_D$ = 14.4 and 13.7 h respectively. The values of the duplication time were in good agreement with that of the control (Fig. 3a). From these results, it could be concluded that the PAA$_{2K}$-coated nanoparticles did not affect significantly the cell proliferation over the 48 h that lasted the experiments. When the cells were exposed to citrate-coated nanoparticles (empty symbols in Figs. 3), the population of adherent fibroblasts started first to decrease by 10 % for Cit–CeO$_2$ and by 30 % for Cit–γ-Fe$_2$O$_3$, before rising again. The data for the iron oxide could not be monitored over the entire time range because the culture medium became turbid, preventing a correct measure of the cell density. Because of the n(t)-decrease at short time, the adjustment using Eq. 5 could not be completed. In conclusion, we have found that in contrast to PAA$_{2K}$-coated particles, the nanoparticles covered with citrates were able to modify the growth laws of the NIH/3T3 cultures. However, these modifications occurred at short times (t < 10 h), and were not deleterious for the long time (t > 24 h) proliferation.

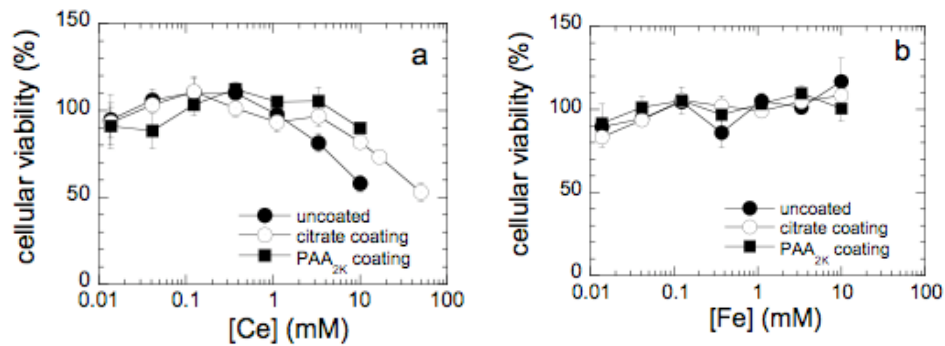

*Figure 4* : *MTT (3-(4,5-dimethylthiazol-2-yl)-2,5-diphenyl tetrazolium bromide) viability assays conducted on NIH/3T3 cells incubated with Cit–CeO$_2$ and PAA$_{2K}$–CeO$_2$ (a) and with Cit–γ-Fe$_2$O$_3$ and PAA$_{2K}$–γ-Fe$_2$O$_3$ (b). The molar concentrations [Ce] and [Fe] were varied from 10 μM to 50 mM and the incubation time was set at 24 h.*

## III.2 – MTT assays

In addition to cell counting, it was important to perform toxicity assays which alone can quantify the cell viability under nanoparticles exposure. MTT (3-(4,5-dimethylthiazol-2-yl)-2,5-diphenyl tetrazolium bromide) viability assays [45] were conducted on NIH/3T3 cells for molar concentrations [Ce] and [Fe] varying from 10 μM to 50 mM and incubation time of 24 h. In terms of weight concentration, the largest value investigated ([Ce] = 50 mM) corresponded to 0.8 wt. %, or equivalently 8 mg mL$^{-1}$. The validity of the present MTT technique was assessed using increasing amounts of DMSO as toxic agent. With DMSO, the decrease of the cell population was observed above 10 mM.

Figs. 4 display the cellular viability for these two particles as a function of the molar concentrations [Ce] and [Fe]. Uncoated, citrate-coated and $PAA_{2K}$-coated particles were surveyed. The 24 h incubation assays (Fig. 4a) show a decrease of the viability at 3 and 10 mM for the uncoated and citrated nanoceria, respectively. For $PAA_{2K}$–$CeO_2$, as well as for the three iron oxide samples (Fig. 4b), the viability remained at a 100 % level within the experimental accuracy. These findings indicate a normal mitochondrial activity for most of the cultures tested. As far as iron oxide is concerned, our results are in good agreement with earlier reports from the literature which showed that both crystalline forms of iron oxide, maghemite $\gamma$-$Fe_2O_3$ and magnetite $Fe_3O_4$ were found biocompatible with respect to cell cultures [5, 7]. Few groups however have noticed positive toxicity for these materials, as for instance Brunner *et al.* who reported acute cell death for flame synthesized iron oxide particles [9]. Concerning cerium oxide, toxicity assays are far less numerous than those performed on iron oxide and some of the toxicity data remained yet controversial. Schubert *et al.* showed that nanoparticles composed of cerium oxide were biocompatible *in vitro* and moreover that these nanomaterials were able to protect HT22 nerve cells from oxidative stress [10]. In contrast, Park and coworkers reported that the exposure of cultured epithelial cells (BEAS-2B) to $CeO_2$ nanoparticles of different sizes led to cell death and to the production of reactive oxygen species (ROS) [13, 46]. As mentioned in earlier toxicity testing [8], these differences might be related to the protective coating adsorbed or tethered on the particle surfaces, as well as to the fate of these particles in the cellular medium. In the present work, we have shown that even at high dose the viability of fibroblasts exposed to nanoceria remains elevated and that the $PAA_{2K}$–coating was playing a crucial role.

## III.3 – Particle Uptake

The uptake of nanoparticles by the cells was monitored by UV-Visible spectrometry. Aliquots of the supernatants in contact with the cells were collected at regular time intervals and analyzed with respect to their oxide concentration. Fig. 5a and 5b display $c(t)/c_0$-data for experiments performed at initial concentration $c_0$ (10 mM) for Cit–$CeO_2$, $PAA_{2K}$–$CeO_2$, Cit–$\gamma$-$Fe_2O_3$ and $PAA_{2K}$–$\gamma$-$Fe_2O_3$. The data obtained with $PAA_{2K}$–$CeO_2$ were found constant within the experimental uncertainty whereas those for Cit–$CeO_2$ exhibited a slight decrease after 24 h. For iron oxide, the $PAA_{2K}$–$\gamma$-$Fe_2O_3$ nanoparticles in the supernatant remained at $c_0$ over the full time range as for the nanoceria, whereas that of Cit–$\gamma$-$Fe_2O_3$ decreased logarithmically by 20 % after 24 h. In the following, we assume that the nanoparticles not present in the supernatants, and thus not detected by spectrometry were either adsorbed on the cellular membranes or taken up by the cells [15, 19]. This assumption allowed us to evaluate the mass of metallic atoms, $M_{Ce}$ and $M_{Fe}$ internalized or adsorbed by the cells as a function of the time. Thanks to the proliferation data of Figs. 3, $M_{Ce}$ and $M_{Fe}$ were normalized with respect to the total number of cells present at a given time, resulting in masses expressed in picogram of cerium or iron per cell [19].

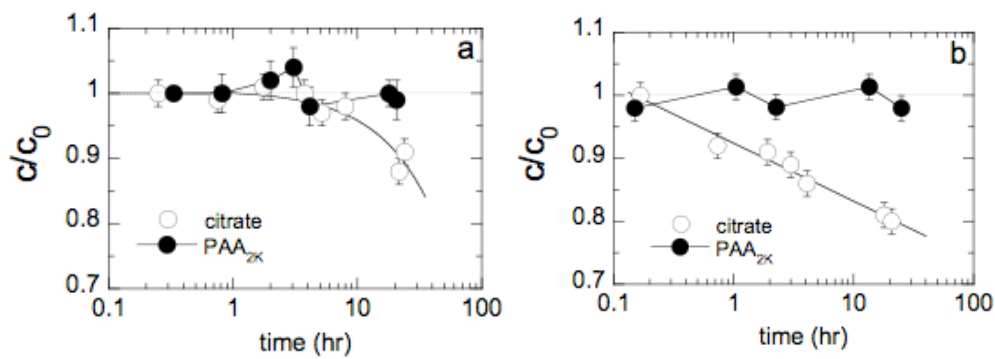

*Figure 5 : Concentrations of cerium and iron oxide nanoparticles in the supernatants of cell culture. The measurements were carried out by UV-Visible spectrometry using absorbance calibration curves for $CeO_2$ and $\gamma$-$Fe_2O_3$ displayed in Supporting Information.*

Fig. 6 compares the mass of iron $M_{Fe}$ incorporated or adsorbed per cell for Cit–$\gamma$-$Fe_2O_3$ and $PAA_{2K}$–$\gamma$-$Fe_2O_3$. As already anticipated from the data in Fig. 5b, $M_{Fe}$ was found to be unchanged for the polymer coated particles whereas it increased logarithmically with time for the citrate-coated particles. After a 24 h incubation, the mass per cell reached a value of 250 pg, a result that compared well to those of the literature on human fibroblasts [19]. The temporal evolution of $M_{Ce}$ exhibited a weaker variation, with a value culminating at 150 pg/cell at 24 h (Fig. SI.5 in Supporting Information).

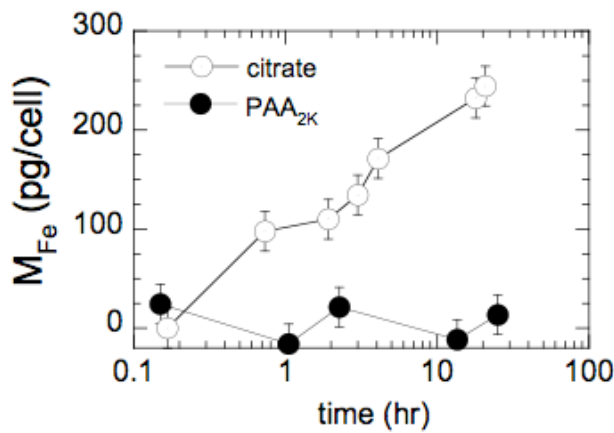

*Figure 6 : Mass $M_{Fe}$ of iron incorporated by the NIH/3T3 cells for experiments conducted with Cit–$\gamma$-$Fe_2O_3$ (empty symbols) and $PAA_{2K}$–$\gamma$-$Fe_2O_3$ (close symbols).*

Figs. 7 displays centrifugation pellets of NIH/3T3 cells that were incubated with Cit–γ-Fe$_2$O$_3$ and PAA$_{2K}$–γ-Fe$_2$O$_3$ nanoparticles for 24 h. After incubation, the cultures were washed thoroughly using PBS in order to remove particles loosely adsorbed onto the cellular membranes. The surface of the T25-flasks were then mechanically scraped and the cell suspensions were centrifuged in Eppendorf vials (5 mn, 1200 rpm, Eppendorf centrifuge 5804R). The exposure of cells to Cit–γ-Fe$_2$O$_3$ resulted in the significant darkening of the centrifugation pellet (Fig. 7b) as compared to the control (Fig. 7a) or to the PAA$_{2K}$–γ-Fe$_2$O$_3$ (Fig. 7c) treated cell line. The data in Figs. 7 data confirmed the previous findings, namely that both anionically coated nanoparticles exhibited very different uptake behaviors against NIH/3T3 cells.

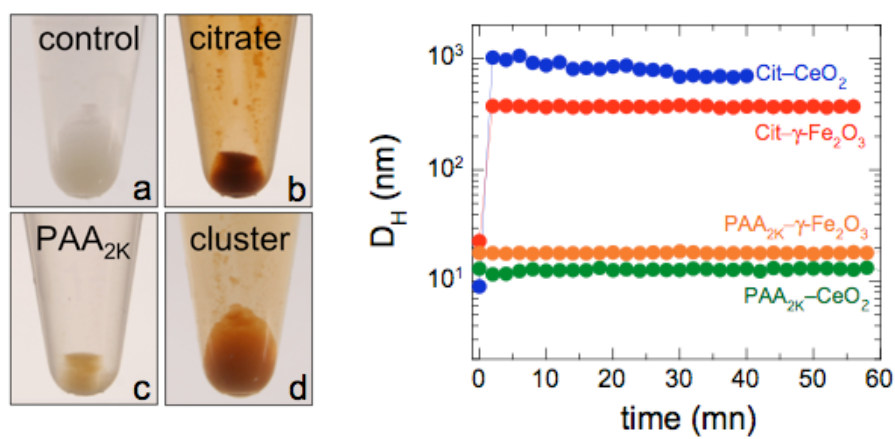

*Figure 7 :* Centrifugation pellets of NIH/3T3 cells that were grown without nanoparticle added (a) and of cells that were incubated with Cit–γ-Fe$_2$O$_3$ (b) and PAA$_{2K}$–γ-Fe$_2$O$_3$ nanoparticles for 24 h (c). In d), the cells were exposed to already preformed 200 nm clusters made of the same iron oxide nanoparticles. A TEM image of these cluster these clusters is provided in Fig. 9.
*Figure 8 :* Time evolutions of the hydrodynamic diameter of Cit–CeO$_2$, PAA$_{2K}$–CeO$_2$, Cit–γ-Fe$_2$O$_3$ and PAA$_{2K}$–γ-Fe$_2$O$_3$ dispersions ([Ce,Fe] = 10 mM) in DMEM supplemented with calf serum and antibiotics. The steep increase in $D_H$ at short times was interpreted as an indication of the destabilization of the dispersion. For the PAA$_{2K}$-coated particles, $D_H$ remained constant over time periods longer than weeks.

## III.4 – Stability in cellular media

From the results of Figs. 3, 6 and 7 it may seem surprising that the NIH/3T3 cells behaved differently with respect to particles which have the same charges at their surfaces. Cell membranes are indeed known to be negatively charged in average and therefore these membranes should exert a net electrostatic repulsion towards surrounding diffusing particles of the same charges. To understand the differences in uptake between citrate and PAA$_{2K}$-coated particles, the colloidal stability of the particles in various solvents, including brines, buffers and cellular growth media was recently put under scrutiny [36]. Here, we underscore the results obtained when the particles were dispersed in the complete culture medium, that is containing the Dulbecco's Modified Eagle's Medium supplemented with 10% fetal bovine serum and 1% penicillin/streptomycin. Fig. 8 displays the time dependence of the hydrodynamic diameters for Cit–$CeO_2$, $PAA_{2K}$–$CeO_2$, Cit–$\gamma$-$Fe_2O_3$ and $PAA_{2K}$–$\gamma$-$Fe_2O_3$, each of the particles having being diluted at t = 0 into the growth medium. Concentrations in the medium were {[Ce,Fe]} = 10 mM for the four specimens. For both citrated particles, at the mixing, $D_H$ exhibited a steep increase, from $D_H$ = 9 and 23 nm to 400 and 1000 nm, respectively. The diameter evolved further and showed a slight decrease at longer time for Cit–$CeO_2$. At longer times (> 1 h), the large aggregates sedimented at the bottom of the test tubes, resulting in the reduction of both scattering intensity and diameter. In contrast, the hydrodynamic diameters for the $PAA_{2K}$-coated nanoparticles remained unchanged, at $D_H$ = 13 and 19 nm respectively. These findings indicate that Cit–$CeO_2$ and Cit–$\gamma$-$Fe_2O_3$ nanoparticles were destabilized at the contact of the culture medium. In Chanteau *et al.* [36], we also demonstrated that this destabilization did not depend on the concentration since it occurred at {[Ce,Fe]} = 0.1 and 1 mM. From the results of Fig. 8, we anticipate that the pronounced uptake exhibited by Cit–$\gamma$-$Fe_2O_3$ could be related to the destabilization of the initially dispersed nanoparticles and their accumulation by gravity in the vicinity of the cell membranes. Obviously, for the $PAA_{2K}$-coated particles, sedimentation did not take place and uptake resulted only by diffusion toward the cells [6].

In order to conclusively prove that the enhanced uptake was due to sedimentation of colloidally unstable particles, the NIH/3T3 fibroblasts were submitted to already formed nanoparticle clusters. We designed recently a novel protocol to aggregate particles into clusters or into rods with submicronic and micronic dimensions [47, 48]. Fig. 9 shows a TEM image of 7 nm $\gamma$-$Fe_2O_3$ clusters obtained by this technique. The aggregates of average diameter 180 nm and polydispersity 0.20 were described as latex-type composite colloids with a high load of magnetic particles. Assuming a volume fraction of 0.25 inside the large spheres [47], we estimated that a 200 nm aggregate was built from ~ 6000 particles. In terms of hydrodynamic sizes, these aggregates compared well with those obtained by the destabilization of the Cit–$\gamma$-$Fe_2O_3$ particles (Fig. 8). Because of the density of iron oxide ($\rho$ = 5100 kg m$^{-3}$), these clusters were also found to sediment readily with time. Exposing iron oxide clusters to fibroblasts resulted again in an enhanced uptake, as illustrated in Fig. 7d. The centrifugation pellet of cells incubated by $\gamma$-$Fe_2O_3$ clusters displayed again a rusty color, indicating an internalization or an adsorption of iron oxide similar to that of the citrated particles (Fig. 7b). Note that the MTT assays using 180 nm spherical clusters were conducted on NIH/3T3 cells and revealed a cell survival rate of 100 % with respect to the control (see Supporting Information). In conclusion, we evidenced that the uptake of nanomaterials depends primarily on the coating of the particles and on its ability to preserve the colloidal nature of the dispersions.

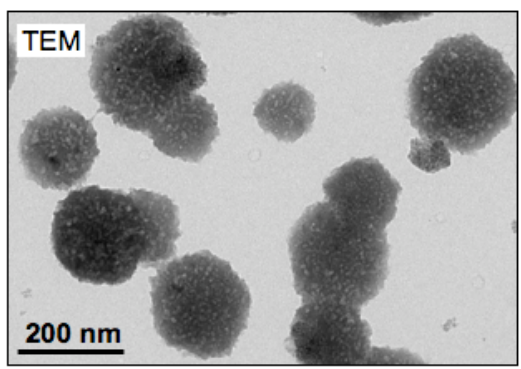

*Figure 9 :* Transmission electron microscopy images (60000×) of spherical aggregates obtained by dialyzing salted dispersions of cationic polymers and anionic γ-Fe$_2$O$_3$ particles [47]. The clusters have an average diameter of 180 nm and a polydispersity of 0.20.

# IV – Conclusions

In this work, the toxicity and uptake of the cerium and iron oxide nanoparticles by NIH/3T3 fibroblasts were investigated. The proliferative properties of the cells and their viability in presence of engineered nanomaterials were evaluated by *i)* transmission optical microscopy to determine the growth laws of the cell populations, *ii)* MTT assays as a function of the metal dose and *iii)* UV-Visible spectrometry for the estimation of the particles uptake. Both particles were synthesized using "soft chemistry" routes which provided nanocolloids of different nature and crystallography, but with the same physico-chemical characteristics. This approach provided us a unique opportunity to compare the influence of the nature of the inorganic materials with respect to that of the coating.

In terms of toxicity, it was shown that even at high dose, most particles were biocompatible as exposed cells remained 100% viable relative to control. Only the bare and the citrate-coated nanoceria exhibited a slight decrease of the mitochondrial activity for cerium concentration above 5 mM (equivalent to 0.8 g L$^{-1}$). These results were found to be in good agreement with those of the literature [5, 7, 49]. In terms of uptake, it was demonstrated that the citrated particles could be internalized/adsorbed by the cells in large amount. The mass of iron incorporated to NIH/3T3 cells was estimated at 250 pg per cell after a 24 h incubation by Cit–γ-Fe$_2$O$_3$ particles. This value was slightly less for cerium. In contrast, the PAA$_{2K}$-coated nanoparticles were taken up at a much lower level, that is below 30 pg/cell. The stronger uptake shown by Cit–γ-Fe$_2$O$_3$ could be related to the destabilization of the initially dispersed nanoparticles in the cell culture medium and their sedimentation near by the surfaces of the cells. For the PAA$_{2K}$-coated particles, either for the cerium and the iron oxides, the polymer coating ensure a long term (> year) stability even in physiological conditions, sedimentation did not take place and uptake resulted only by diffusion and single adsorption on the cell membranes. These results also suggest that anionically charged polymers represent an effective alternative to conventional coating agents.

**Acknowledgement**
We thank Dietrich Averbeck, Armelle Baeza-Squiban, Jean-Paul Chapel, Marco El Rawi, Jérôme Fresnais, Antje Neeb, Régine Perzynski, Sandra Schneider, and Carsten Weiss for numerous and fruitful discussions. The Rhodia R&D research laboratory (Aubervilliers, France) are acknowledged for providing us with the nanoceria. Aude Michel (PECSA, Université Pierre et Marie Curie, Paris, France) is kindly acknowledged for the TEM experiments. This research was supported in part by Rhodia (France), by the Agence Nationale de la Recherche under the contract BLAN07-3_206866, by the European Community through the project : "NANO3T—Biofunctionalized Metal and Magnetic Nanoparticles for Targeted Tumor Therapy", project number 214137 (FP7-NMP-2007-SMALL-1) and by the Région Ile-de-France in the DIM framework related to Health, Environnement and Toxicology (SEnT).

doi:10.1088/0957-4484/21/14/145103     Nanotechnology 21 (2010) 145103
Published 16 March 2010[24] Hafelli UO, Riffle JS, Harris-Shekhawat L, Carmichael-Baranauskas A, Mark F, Dailey JP, et al., editors. Cell Uptake and in Vitro Toxicity of Magnetic Nanoparticles Suitable for Drug Delivery. NanoMedicine Summit on Nanoparticles for Imaging, Diagnosis, and Therapeutics; 2008 Sep 25; Cleveland, OH: Amer Chemical Soc.

[25] Tromsdorf UI, Bruns OT, Salmen SC, Beisiegel U, Weller H. A Highly Effective, Nontoxic T1 MR Contrast Agent Based on Ultrasmall PEGylated Iron Oxide Nanoparticles. Nano Letters. 2009;9(12):4434-40.

[26] Chastellain M, Petri A, Hofmann H. Particle size investigations of a multistep synthesis of PVA coated superparamagnetic nanoparticles. J Colloid Interface Sci. 2004;278(2):353 - 60.

[27] Petri-Fink A, Chastellain M, Juillerat-Jeanneret L, Ferrari A, Hofmann H. Development of functionalized superparamagnetic iron oxide nanoparticles for interaction with human cancer cells. Biomaterials. 2005;26(15):2685-94.

[28] Petri-Fink A, Steitz B, Finka A, Salaklang J, Hofmann H. Effect of cell media on polymer coated superparamagnetic iron oxide nanoparticles (SPIONs): Colloidal stability, cytotoxicity, and cellular uptake studies. European Journal of Pharmaceutics and Biopharmaceutics. 2008;68(1):129-37.

[29] Berret J-F. Stoichiometry of electrostatic complexes determined by light scattering. Macromolecules. 2007;40(12):4260-6.

[30] Qi L, Sehgal A, Castaing JC, Chapel JP, Fresnais J, Berret JF, et al. Redispersible hybrid nanopowders: Cerium oxide nanoparticle complexes with phosphonated-PEG oligomers. Acs Nano. 2008;2(5):879-88.

[31] Lin C-AJ, Sperling RA, Li JK, Yang T-Y, Li P-Y, Zanella M, et al. Design of an Amphiphilic Polymer for Nanoparticle Coating and Functionalization13. Small. 2008;4(3):334-41.

[32] Limbach LK, Bereiter R, Mueller E, Krebs R, Gaelli R, Stark WJ. Removal of oxide nanoparticles in a model wastewater treatment plant: Influence of agglomeration and surfactants on clearing efficiency. Environmental Science & Technology. 2008;42(15):5828-33.

[33] Roca AG, Costo R, Rebolledo AF, Veintemillas-Verdaguer S, Tartaj P, Gonzalez-Carreno T, et al. Progress in the preparation of magnetic nanoparticles for applications in biomedicine. Journal of Physics D: Applied Physics. 2009;(22):224002.

[34] Sehgal A, Lalatonne Y, Berret J-F, Morvan M. Precipitation-redispersion of cerium oxide nanoparticles with poly(acrylic acid): Toward stable dispersions. Langmuir. 2005;21(20):9359-64.

[35] Berret J-F, Sandre O, Mauger A. Size distribution of superparamagnetic particles determined by magnetic sedimentation. Langmuir. 2007;23(6):2993-9.

[36] Chanteau B, Fresnais J, Berret JF. Electrosteric Enhanced Stability of Functional Sub-10 nm Cerium and Iron Oxide Particles in Cell Culture Medium. Langmuir. 2009;25(16):9064-70.

[37] Bacri JC, Perzynski R, Salin D, Cabuil V, Massart R. Magnetic colloidal properties of ionic ferrofluids. Journal of Magnetism and Magnetic Materials. 1986;62(1):36-46.

[38] Dubois E, Cabuil V, Boue F, Perzynski R. Structural analogy between aqueous and oily magnetic fluids. J Chem Phys. 1999;111(15):7147 - 60.

[39] Spalla O, Cabane B. Growth of Colloidal Aggregates through Polymer Bridging. Colloid Polym Sci. 1993;271:357 - 71.

[40] Biggs S, Healy TW. Electrosteric Stabilization of Coloidal Zirconia with Low Molecular Weight Poly(Acrylic Acid). J Chem Soc Faraday Trans. 1994;90:3415 - 21.
20